# Abnormal filtering property based on the divergence of the impedance in ladder-shape network consisting of inductors and capacitors

Guoan Zheng[1], Fang Cai[2], Mingwu Gao[2]

1. Department of Optical Engineering Zhejiang University ; 2. Department of Electronic Engineering Zhejiang University, Hangzhou, China, 310027

**Abstract**: The total impedance of a ladder-shape network consisting of inductors and capacitors does not converge to a certain value when the steps of the network increased. In this paper, we analyze this effect in frequency domain. We find that in some band the impedance converge to a limit value while in other band it doesn't. Based on this property in frequency, we propose a filter that exhibits excellent performance both in amplitude and phase response. As a validation of our result, the simulation of this filter was carried out on the EDA software Multisim with respect to a practical circuit.

## 1. Introduction

An n steps ladder-shape network is shown in Fig.1 (a). In order to calculate the total resistor $R_n$ of the network, we use the equation based on the equivalent circuit shown in Fig.1(b):

$$R_n = r_2 + \frac{r_1 R_{n-1}}{r_1 + R_{n-1}} \quad (1)$$

If $n$ is large enough maybe it is justifiable to believe that:

$$R_n = R_{n-1} (n \to \infty) \quad (2)$$

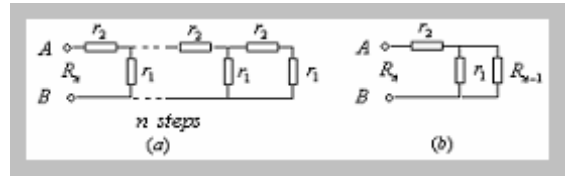

Fig.1 (a) a n steps ladder-shape network
(b) equivalent circuit for Fig.1(a)

From Eq.1 and Eq.2, we get:

$$R_n = \frac{r_2 + \sqrt{r_2^2 + 4r_1 r_2}}{2} \quad (n \to \infty) \quad (3)$$

If the ladder-shape network consists of inductors and capacitors, as shown in Fig.2, we simply substitute in Eq.3 $iwL$ for $r_2$, $1/i\omega C$ for $r_1$ and $Z_n$ for $R_n$, with the result:

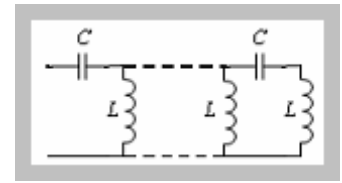

Fig.2 ladder-shape network consisting of inductors and capacitors

$$Z_n = \frac{1 + \sqrt{1 - 4\omega^2 LC}}{2i\omega C} \quad (n \to \infty) \quad (4)$$

As pointed out in Ref.1, this result is not true when $1 - 4\omega^2 LC < 0$, because Eq.4 has a nonzero real part, which means energy loss in the network that only consisting of inductors and capacitors. Therefore Eq.2 is not true and when the order of ladder-shape network increases, the impedance does not converge to a limit value.

In this paper we study this property of divergence in frequency domain. In section2 we show that when $1 - 4\omega^2 LC > 0$, the impedance $Z_n$ do converge to a limit value. Here comes to an interesting point that it does exist a critical frequency $f_c$, below and beyond this $f_c$, the behavior of the ladder-shape network is totally different. When $f < f_c$, the impedance of ladder-shape network does converge to a limit value whereas it doesn't in the case of $f > f_c$. Based on this property in frequency, a filter is proposed in section3. We analyze the theoretical model with software Matlab then construct the practical circuit and do simulation on the EDA software Multisim7 as a validation of our result. Finally we get our conclusion in Part 4.

## 2. Special behavior of total impedance in frequency domain

From Eq.4, let $1-4\omega^2 LC = 0$ we get a critical frequency $\omega_c = 1/2\sqrt{LC}$ ($f_c = \omega_c/2\pi$). In order to study the behavior of the total impedance beyond and below this critical frequency $f_c$, we make the total impedance of an n step ladder-shape network (represented by $Z_n$) as a function of the frequency $f$, as shown in Fig.3. Different curves in Fig.3 represent the impedance of network with step 9, 10 and 11 respectively. The calculation is based on $C = 1nF$ and $L = 1mH$. From the figure depicted by Matlab, we obtain the critical frequency 80kHz, which complies with theoretical value (79.6kHz) very well. For $f < f_c$, three curves approximately superpose each other. However, for $f > f_c$, the step of network has substantial effect on the impedance hence three curves do not superpose each other any more. In sum, the property of total impedance of the ladder-shape network with capacitors and inductors below and beyond the critical frequency $f_c$ behaves totally different. And this characteristic is the theoretic foundation of the filter we shall mention in the following.

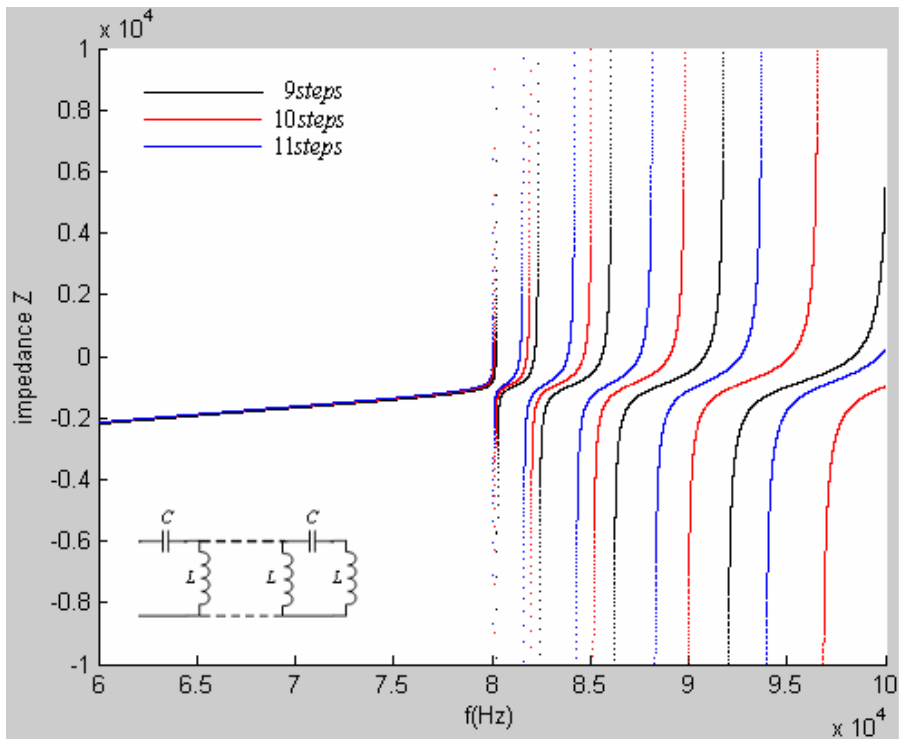

Fig.3 Curves of impedances of ladder-shape networks with different steps varying with frequency $f$. The calculation is based on $C = 1nF$ and $L = 1mH$. The critical point we get in the figure matches the theoretical deduction perfectly. For $f < f_c$, the difference of network order contributes very little to total impedance, but for $f > f_c$, the situation is totally opposite, as you may find out clearly from the curve.

As shown in Fig.4, ordinate represents the amplitude of the equivalent impedance $Z_{eq} = Z_9 Z_{10}/(Z_9 - Z_{10})$ of $Z_{10}$ and -$Z_9$ connecting in parallel while abscissa represents frequency. The subscript of $Z$ means the steps of the ladder-shape network. The calculation is based on $C = 1nF$ and $L = 1mH$. The abnormal variation trend comes from the fact that the value of $Z_{10} - Z_9$ is close to zero when $f < f_c$, while $Z_{10} - Z_9$ is a comparatively large value when $f > f_c$ as shown in Fig.3. In addition, we can utilize negative impedance converter[3] to realize the subtraction of two networks with different steps.

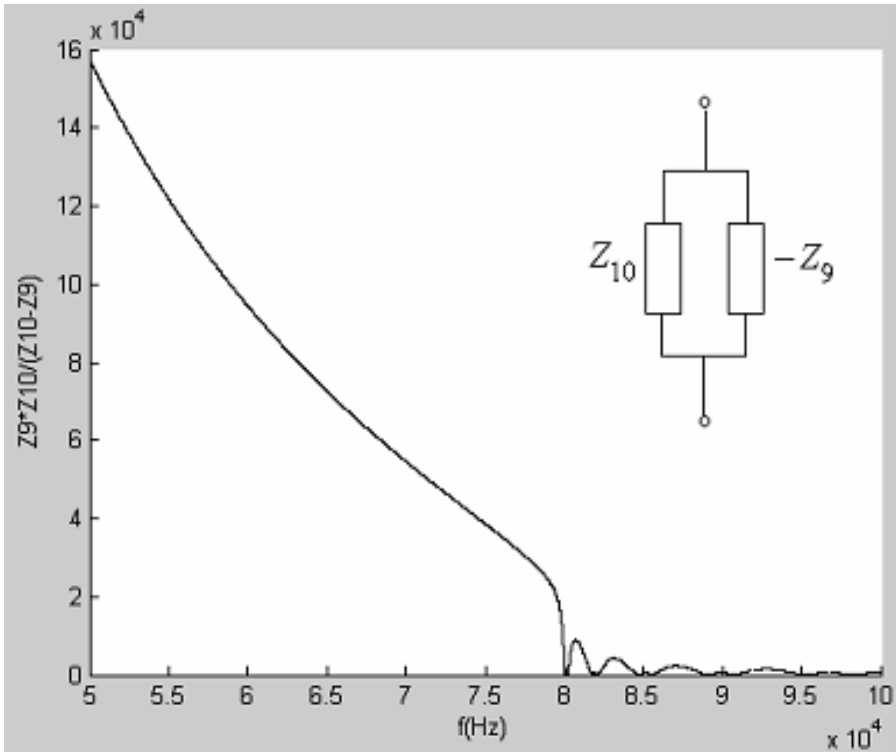

Fig.4 Ordinate represents the equivalent impedance $Z_{eq} = Z_9 Z_{10}/(Z_9 - Z_{10})$, while abscissa represents frequency. When $f < f_c$, the value of $Z_{10} - Z_9$ is close to zero and therefore $Z_{eq}$ is a large value. However, when $f > f_c$ the value of $Z_{10} - Z_9$ is a comparatively large value and therefore the value of $Z_{eq}$ is small. Curve here also verifies the result of critical point along with Fig.3.

## 3. Filter design based on special behavior of total impedance in frequency domain

Taken advantage of this special behavior, a high-pass filter is proposed here. Its model circuit is shown in Fig.5, The subscript of $Z$ means the steps of the ladder-shape network.. Still, $C = 1nF$ and $L = 1mH$. And $-Z_9$ denote the equivalent impedance of $Z_9$ after transformed by negative impedance converter[7]. Resistor $R$ is chosen to a large value($50kOhm$ in our design). $Z_{10}$ and $-Z_9$ are connected in parallel. The equivalent impedance is $Z_{eq} = Z_9 Z_{10}/(Z_9 - Z_{10})$ and its variation trend with frequency is shown in Fig.4. Note that the value of resistor $R$ should be carefully selected so that it is significantly greater than $Z_{eq}$ when $f > f_c$ and much less than $Z_{eq}$ when $f < f_c$. Since $V_{out} = V_{in} R/(Z_{eq} + R)$, when $f > f_c$, $V_{out}$ is close to $V_{in}$ and when $f < f_c$, $V_{out}$ is close to 0. Therefore this system can function as a high-pass filter.

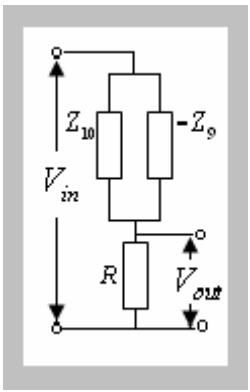

Fig.5 The model of the proposed filter. $Z_{10}$ and $-Z_9$ are connected in parallel, their equivalent value is $Z_{eq} = Z_9 Z_{10}/(Z_9 - Z_{10})$. $Z_{eq}$ and R are connected in series. When $f < f_c$, since the value of $Z_9 - Z_{10}$ is small, $Z_{eq}$ is much greater than R and therefore $V_{out}$ is close to $V_{in}$; when $f > f_c$, since the value of $Z_9 - Z_{10}$ is comparatively large, $Z_{eq}$ is much less than $R$ and therefore $V_{out}$ is close to zero.

Based on this theoretical analysis, we simulate the result using Multisim 7. The simulated circuit is shown in Fig.6. Amplitude and phase responses to frequency are shown in Fig.7. The amplitude response illustrates the high-pass property of the designed filter. From the trend of phase response to frequency we can see that when in the pass-band $f > f_c$, phase response shows good linearity property[6], group delay $-d\varphi/d\omega$ is close to zero. When $f < f_c$, phase response is not within our consideration because in stop-band the amplitude is attenuated greatly. Fig.7 shows these properties of the filter.

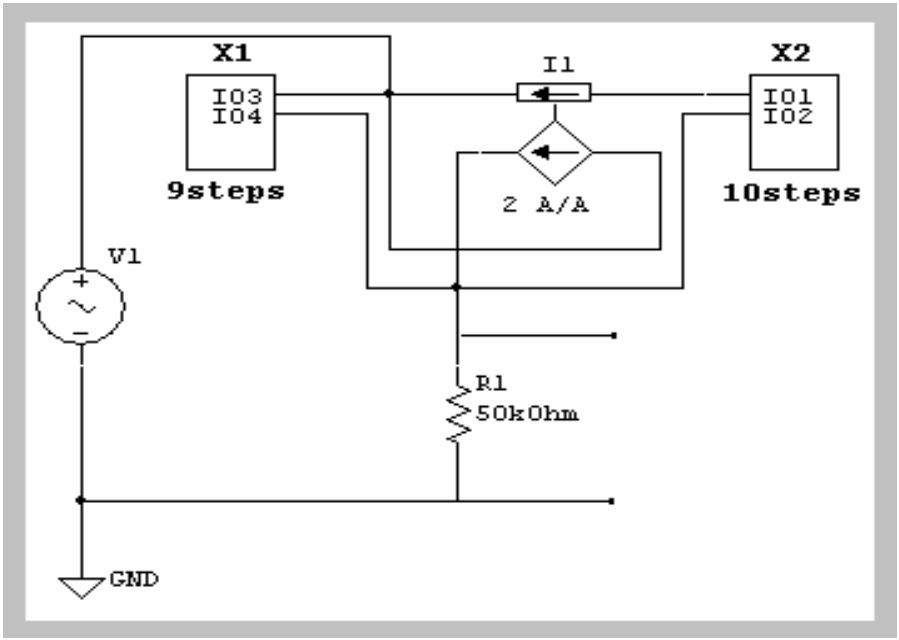

Fig. 6 The simulated circuit of the designed filter in Multisim7. $X_1 = Z_9$ and $X_2 = Z_{10}$. The subscript of $Z$ means the steps of the ladder-shape network.. I1 is a current control current source which is used as a negative impedance converter here. Through CCCS, $X_2$ is able to be converted into $-X_2$. $X_1$ and $-X_2$ are connected in parallel. After that, they are connected with $R_1$ =50kOhm in series just as the theoretical model shown in Fig.5.

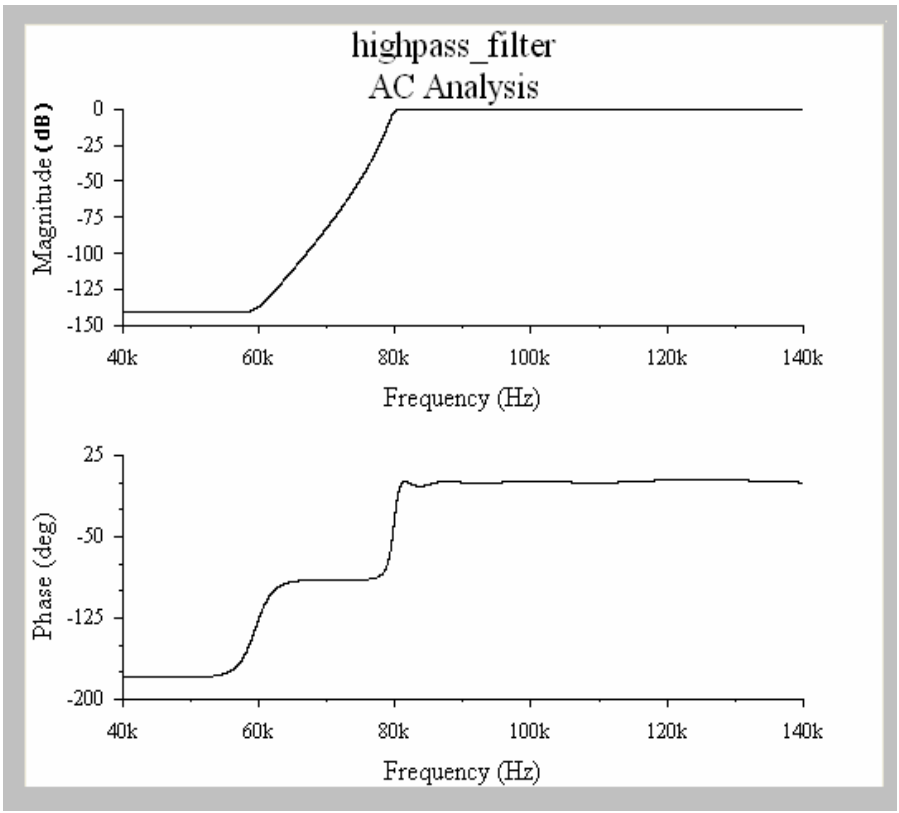

Fig.7 The amplitude and phase responses to frequency for the practical circuit. The amplitude response illustrates the high-pass property of our filter. The phase response shows that the group delay is close to zero, which means that the real-time property is quite good in our design. The linearity of phase response in the pass-band is also shown in the figure.

### 4. Conclusion

We have proposed and demonstrated an LC ladder-shape filter based on the special behavior of total impedance in frequency domain. Since linearity plays an important role in the property of a filter, our filter has a quite perfect linearity in the pass-band. The group delay $-d\varphi/d\omega$ is close to zero in the pass-band which means the real-time property is quite good in our design.

**Address: Mixed class of CKC honors college, Yuquan campus of Zhejiang University ,Hangzhou , China, 310027**
**Linkman: Zheng Guo An (grade of 2003)**
**E-mail :zhengguoan1984@yahoo.com.cn**
          zhengguoan1984@163.com